\documentstyle[aps,multicol,epsf,amsbsy]{revtex}

\begin{document}

\draft

\title{Universality in the off-equilibrium critical dynamics of the 
$3d$ diluted Ising model}

\author{G. Parisi\thanks{e-mail: {\tt
giorgio.parisi@roma1.infn.it}}$^{,1}$,
F. Ricci-Tersenghi\thanks{e-mail: {\tt
federico.ricci@roma1.infn.it}}$^{,1}$ and
J. J. Ruiz-Lorenzo\thanks{e-mail: {\tt
ruiz@lattice.fis.ucm.es}}$^{,2}$}

\address{$^1$ Dipartimento di Fisica,
Universit\'a di Roma ``La Sapienza'',
Piazzale Aldo Moro 2, 00185 Roma, Italy.\\
$^2$ Departamento de F\'{\i}sica Te\'orica I, 
Universidad Complutense de Madrid,
Ciudad Universitaria, 28040 Madrid, Spain.}

\date{\today}
\maketitle

\begin{abstract}
We study the off-equilibrium critical dynamics of the three
dimensional diluted Ising model.  We compute the dynamical critical
exponent $z$ and we show that it is independent of the dilution only
when we take into account the scaling-corrections to the dynamics.
Finally we will compare our results with the experimental data.
\end{abstract}
\pacs{PACS numbers: 05.50.+q, 75.10.Nr, 75.40.Mg}

\begin{multicols}{2}
\narrowtext

The issue of Universality in disordered systems is a controversial and
interesting subject.

Very often in the past it has been argued that critical exponents
change with the strength of the disorder~\cite{REVIEW_HEUER}.  While,
on a deeper analysis, it has turned out that those exponents were
``effective'' ones, {\it i.e.}\ they are affected by strong scaling
corrections.  So, when one studies the critical behavior of a
disordered system it is mandatory to control the leading
correction-to-scaling in order to avoid these effects that could
modify the dilution-independent values of the critical exponents.  For
instance, in Ref.~\cite{DILU3D} the equilibrium critical behavior of
the three dimensional diluted Ising model was studied. The authors
showed that taking into account the corrections-to-scaling it was
possible to show that the static critical exponents ({\it e.g.}\ $\nu$
and $\eta$) and cumulants were dilution-independent.  These numerical
facts supports the (static) perturbative renormalization group
picture: all the points of the critical line (with $p<1$) belong to
the same Universality class (with critical exponents given by the
random fixed point)~\cite{AHARONY}.  Their final values of the
exponents~\cite{DILU3D} were in very good agreement with the
experimental figures (see below).

We will show that an analogous effect also happens in the
off-equilibrium dynamics of the diluted ferromagnetic model and we
will take it into account in our data analysis, in order to get the
best estimate of the critical dynamical exponent.

The critical dynamics of the diluted Ising model has been studied
experimentally in Ref.~\cite{BELANGER} using neutron spin-echo
inelastic scattering on samples of ${\rm Fe}_{0.46}{\rm Zn}_{0.54}
{\rm F}_2$ (antiferromagnetic diluted model) and has been compared
with the results obtained in pure samples (${\rm Fe} {\rm
F}_2$)~\cite{BELANGER}. For the pure model a dynamical critical
exponent $z=2.1(1)$ was found (in good agreement with the theoretical
predictions based on one-loop perturbative renormalization group
(PRG)~\cite{MA}) whereas in the diluted case the exponent $z=1.7(2)$
was computed (three standard deviations away of the analytical
prediction based on (one-loop) PRG that provides $z\simeq
2.34$~\cite{GRINSTEIN}). Furthermore, the dynamical exponent was
computed in the framework of the PRG up to two loops and it was
obtained $z=2.237$~\cite{PRUDNIKOV} and $z=2.180$~\cite{JANSSEN} (the
experimental value is at 2.5 standard deviation of the two loops
analytical result).

In the experiment~\cite{BELANGER} were measured critical amplitudes
100 times smaller than those computed in the pure case. It is clear
that a more precise experiment on this issue will be welcome. We
should point out that the critical dynamics of a diluted
antiferromagnet is the same as of a diluted ferromagnet.

A numerical study of the on-equilibrium dynamics in diluted systems
was performed in 1993 by Heuer~\cite{HEUER}.  He measured the
equilibrium autocorrelation functions for different concentrations and
lattice sizes.  The autocorrelation time ($\tau$) depends on the
lattice size ($L$) via the formula $\tau \propto L^z$ (neglecting
scaling corrections).  He found that all the data, for concentrations
not too close to 1, were compatible, for large $L$, with the
assumption of a single dynamical exponent, different from the one of
the pure fixed point and similar to the analytical estimate of
Ref.~\cite{GRINSTEIN} ($z\simeq 2.3$). The final value reported by
Heuer was $z=2.4(1)$.

The main goal of this work is to check Universality in the critical
dynamics of diluted models ({\it i.e.}\ whether the dynamical critical
exponent is dilution independent) in the off-equilibrium
regime~\cite{NOTE}.  To do this we monitor scaling corrections in the
same way it was done in the static simulations~\cite{DILU3D}.
Therefore we will also obtain the value of the corrections-to-scaling
exponent for the dynamics. Our motivation to study the off-equilibrium
dynamics instead of the equilibrium one is based on two reasons.  The
more important reason is that the experimental data was obtained in the
off-equilibrium regime and the second one is that (in general) it is
easier to simulate systems in the off-equilibrium regime. Moreover it
will be possible to confront our $z$ computed in the off-equilibrium
regime with that obtained at equilibrium~\cite{HEUER}.

The relevance of the corrections-to-scaling is twofold. The first one
is that the scaling-corrections are very important in the right
determination of the static (equilibrium simulation) critical
exponents~\cite{DILU3D}.  In some models the corrections-to-scaling
change the anomalous dimension of the order of 10 \% (see for example
Ref.~\cite{PERCO4D}).  The second one is that the
correction-to-scaling exponent can be (and it has been) computed in a
real experiment~\cite{OMEGAEXP}.

We have studied the three dimensional diluted Ising model defined on a
cubic lattice of size $L$ and with Hamiltonian
\begin{equation}
{\cal H} = -\sum_{<ij>} \epsilon_i \epsilon_j S_i S_j \quad ,
\end{equation}
where $S_i$ are Ising spin variables, $<\!\!ij\!\!>$ denotes sum over
all the nearest-neighbor pairs and $\epsilon_i$ are uncorrelated
quenched variables which are 1 with probability $p$ and zero
otherwise.

We have measured, at the infinite volume critical point and for
several concentrations $p$, the non-connected susceptibility, defined
by
\begin{equation}
\chi = \frac{1}{L^3} \sum_{ij} \overline{\langle S_i S_j \rangle}
\quad ,
\label{E-chi}
\end{equation}
where the brackets stand for the average over different thermal
histories or initial configurations and the horizontal bar for an
average over the disorder realizations. The indices $i$ and $j$ run
over all the points of the cubic lattice.  In practice we use a large
number of disorder realizations ($N_S=512$) each with a single thermal
history, what amounts to neglect the angular brackets in
Eq.(\ref{E-chi}).  This procedure is safe and does not introduce any
bias.

With the notation of the book of Ma~\cite{MA} we can write, for
instance, the following equation for the response function, under a
transformation of the dynamical renormalization group (RG) with step
$s$,
\begin{eqnarray}
\lefteqn{G(\boldsymbol k, \omega, \boldsymbol \mu) =} \nonumber \\
& s^{2-\eta} G\left(s \boldsymbol k, s^z \omega, \boldsymbol \mu^* \pm
\left(\frac{s}{\xi}\right)^{y_1} \boldsymbol e_1 + O(s^{y_2})\right)
\quad ,
\label{scaling}
\end{eqnarray}
where $\omega$ is the frequency, $\boldsymbol k$ is the wavelength
vector, $z$ is the dynamical critical exponent, by $\boldsymbol \mu$
we denote all the parameters of the Hamiltonian, $\boldsymbol \mu^*$
is the fixed point of the renormalization group transformation, $\xi$
is the static correlation length and finally $y_1$ is the relevant
eigenvalue (equals to $1/\nu$: $y_1$ is the scaling exponent
associated with the reduced temperature), $\boldsymbol e_1$ is its
associate eigenvector and $y_2$ is the greatest irrelevant eigenvalue
($y_2<0$) of the renormalization group transformation (we have assumed
that the system posses only one relevant operator).

Using Eq.(\ref{scaling}) and considering the leading
scaling-corrections for a very large system~\cite{NOTE1} at the
critical temperature, we can write the dependence of the
susceptibility on the Monte Carlo time as
\begin{equation}
\chi(t,T_c(p)) = A(p) t^{\frac{\gamma}{\nu z}} + B(p)
t^{\frac{\gamma}{\nu z}-\frac{w}{z}} \quad ,
\label{sus-tot}
\end{equation}
where $t$ is the Monte Carlo time, $T_c(p)$ is the critical
temperature, $A(p)$ and $B(p)$ are functions that depend only on the
spin concentration, $\gamma$ is the exponent of the static
susceptibility, $\nu$ is the exponent of the static correlation
length, $z$ is the dynamical critical exponent and finally $w \equiv
-y_2$ is the correction-to-scaling exponent. Hereafter we denote $w_d
\equiv w/z$. We recall that $w$ corresponds with the biggest
irrelevant eigenvalue of the RG in the dynamics, in principle $w$ will
be different from the leading correction in the static (that we will
denote by $w_s$)~\cite{MA}.  In addition, an analytical
correction-to-scaling comes from the non singular part of the free
energy and gives us a background to add to Eq.(\ref{sus-tot}). In our
numerical simulations we can neglect this background term ({\it i.e.}\
we will show that $\gamma/(\nu z)-\omega/z\simeq 0.5 \gg 0$).

Moreover, Eq.(\ref{sus-tot}) is valid for times larger than a given
``microscopic'' time and for times (in a finite lattice) less than the
equilibration time (that is finite in a finite lattice).

To study numerically the present issue we have simulated $L=100$
systems for different spin concentrations $p = 1.0, 0.9, 0.8, 0.65,
0.6, 0.5$ and $0.4$ at the critical temperatures reported in
Ref.~\cite{DILU3D}.  The Metropolis dynamics~\cite{METROPOLIS}
provides our local dynamics.  We have checked in all the simulations
that we were in an off-equilibrium situation: for the volumes and
times we have used, the non-connected susceptibility is far from
reaching its equilibrium plateau (in a finite system).  For
completeness we also report the numerical estimate of the critical
exponents for the random fixed point, where all the systems with $p<1$
should converge for large length scales~\cite{DILU3D}:
$\gamma=1.34(1)$, $\nu=0.6837(53)$, $\eta=0.0374(45)$ and
$w_s=0.37(6)$ (Ref.~\cite{HOLOVATCH}, using PRG, provides
$\omega=0.372(5)$ in the massive scheme and $\omega=0.39(4)$ in the
minimal subtraction one).  It is worth noting that experimentally the
best estimate of the susceptibility exponent is
$\gamma=1.33(2)$~\cite{BELANGER2}.

At this point we can recall the one-loop prediction of the PRG for the
$\nu$ and $\eta$ exponents: $\nu=\frac{1}{2} + \frac{1}{4}
\sqrt{\frac{6 \epsilon}{53}}$ and
$\eta=-\epsilon/106$~\cite{AHARONY,KLEINERT}, where $\epsilon=4-d$,
$d$ being the dimensionality of the space.  If we substitute
$\epsilon=1$ we obtain the following (``naive'') estimates:
$\eta=-0.0094$ and $\nu=0.5841$. Obviously the previous naive
estimates are far from the numerical and experimental values of the
critical exponents. This would also imply that even the one-loop PRG
estimate of the dynamical critical exponents will stay far from the
true value.

Notice also that the anomalous dimension exponent ($\eta$) takes near
the same value either at the pure and at the random fixed point. One
can argue that this holds using the arguments provide in
Ref. \cite{CARDY} using a $\epsilon^\prime$-expansion (where $d \equiv
2 + \epsilon^\prime$)~\cite{NOTE3}.  This fact, assuming the naive
dynamical theory (Van Hove theory or conventional theory)~\cite{MA},
implies that the dynamical critical exponent $z = \gamma/\nu = 2-\eta$
is the same for both, diluted and pure system, to first order in
$\epsilon^\prime$. We will show that this is not the case for our
diluted model. The Van Hove theory was used in~\cite{BELANGER} to
interpret the experimental data.

An analytical estimate of the value of the dynamical critical exponent
has been taken from Ref.~\cite{GRINSTEIN} where a dynamical
$\sqrt{\epsilon}$-expansion ($\epsilon \equiv 4-d$) was done:
$z=2+\sqrt{6\epsilon/53}+O(\epsilon)$, that in three dimension becomes
$z\simeq 2.34$, where we have neglected the terms $O(\epsilon)$. We
can recall the two loops computation: $z=2.237$~\cite{PRUDNIKOV} and
$z=2.180$~\cite{JANSSEN}.  One of the results of this work should be
about the reliability of the previous estimates of $z$ (the first and
second term of an $\sqrt \epsilon$-expansion).

With all these ingredients we can analyze our numerical data for the
dynamical non-connected susceptibility and check whether or not the
Universality, based on renormalization group arguments, holds.

\begin{figure}
\epsfxsize=0.95\columnwidth
\epsffile{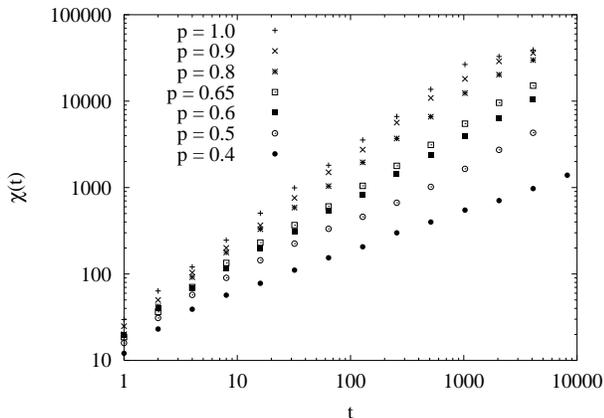}
\caption{The growth of the out-of-equilibrium susceptibility with the
Monte Carlo time, at the critical temperature.  The lattice volume is
always $100^3$ and the spin concentrations are reported in the plot.
The errors are smaller than the symbols.}
\label{fig:raw}
\end{figure}

In the first plot (Fig.~\ref{fig:raw}) we show the numerical data in a
double logarithm scale. The slope gives, neglecting the
corrections-to-scaling, the ratio $\gamma/(\nu z)$. It seems that all
the lines behave in a power law but with different slopes [{\it i.e.}\
different exponents $\gamma/(\nu z)$]. This fact could call for non
universality in this model ({\it i.e.}\ critical exponents vary along
the critical line).  In addition, if we take into account the main
result from the static~\cite{DILU3D} which states that the static
critical exponents ({\it e.g.}\ $\nu$ and $\eta$) do not depend on the
dilution degree, we obtain a dynamical exponent that depends on the
dilution, violating the prediction of the dynamical perturbative
renormalization group~\cite{GRINSTEIN}.  In fact, following the RG
flow (for $p<1$) we should always end at the same random fixed point
and so, for large scales (in time and space) $z$ is not expected to
depend on the dilution degree.

In the previous analysis we have not taken into account the scaling
corrections. However, we are able to monitor the leading
scaling-corrections given by the exponent $w$. We succeeded in fitting
(using the MINUIT routine~\cite{MINUIT}) all our numerical data to
Eq.(\ref{sus-tot}) for $0.5\le p \le 0.8$. We have 10 parameters to
fit : $A(p)$ and $B(p)$ for four dilutions ($p=0.8$, $0.65$, $0.6$,
$0.5$), $\gamma/(\nu z)$ and $\omega/z$, these last exponents assumed
dilution independent.

In this way we have computed the functions $A(p)$ and $B(p)$ in
Eq.(\ref{sus-tot}) and $\gamma/(\nu z)$ and $\omega/z$. By fitting the
data using $t \ge 4$ we have obtained a very good fit (with
$\chi^2/{\rm d.o.f}=33.8/34$, where d.o.f stands for degrees of
freedom) and the following values for the dynamical critical exponent
and the leading dynamical scaling-corrections
\begin{equation}
z=2.62(7) \,\,\, , \,\,\, \omega=0.50(13) ,
\end{equation}
where we have used the value of the static critical exponents
$\gamma=1.34(1)$ and $\nu=0.6837(53)$~\cite{DILU3D}.

In order to check the stability of the previous fit we have tried a
new fit using only times $t\ge 8$. The fit again is very good (with
$\chi^2/{\rm d.o.f}=29.7/30$) and
\begin{equation}
z=2.58(7) \,\,\, , \,\,\, \omega=0.72(16) ,
\end{equation}
Clearly the fit is very stable since both exponents are compatible
inside the error bars (one half standard deviation in $z$ and one standard
deviation in $\omega$).  Therefore we take, as our final values,
$z=2.62(7)$ and $\omega=0.50(13)$.

\begin{figure}
\epsfxsize=0.95\columnwidth
\epsffile{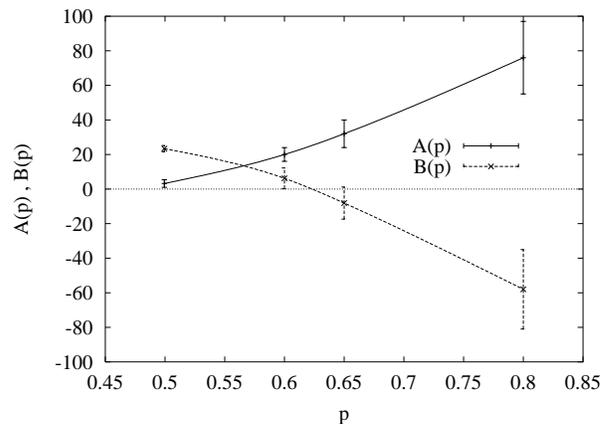}
\caption{The amplitudes defined by Eq.(\protect\ref{sus-tot}) are
smooth functions of the spin concentration.  Where the $B(p)$ crosses
the axis a ``perfect Hamiltonian'' can be defined (see text).}
\label{fig:ab}
\end{figure}

In Fig.~\ref{fig:ab} we show our results for the amplitudes $A(p)$ and
$B(p)$ (using the results of the fit with $t \ge 4$; $t=4$ plays the
role of the microscopic time for this model and algorithm, see the
previous discussion). The main result of these fits is that the
numerical data can be well described using dilution independent
exponent (both dynamical and static), while the value of the dilution
only enters in the non-universal amplitudes, $A(p)$ and $B(p)$. This
fact clearly supports Universality in this model.

From Fig.~\ref{fig:ab} we can compute the value of the dilution in
which there is not (leading) scaling-corrections (one kind of
``perfect Hamiltonian'' for this dynamical problem). For $p \simeq
0.63$ we obtain $B(p) \simeq 0$ and so with this dilution it is
possible to measure dynamical critical exponents [{\it e.g.}\
$\gamma/(z \nu)$ from the growth of the susceptibility, $(d-1/\nu)/(z
\nu)$ from the relaxation of the energy, etc.] neglecting the
underlying (leading) scaling-corrections. This dilution could be a
good starting point in order to monitor the sub-leading
scaling-corrections.

Systems with spin concentrations $p=0.9$ have also been simulated, but
the data from these runs are not been included in the previous
analysis, because they can not be well fitted with the formula of
Eq.(\ref{sus-tot}).  We can explain this fact assuming that for this
dilution the system is in the cross-over region, for the lattice and
times we used.  Also in the static studies a similar effect was found
and only for $p\le0.8$ was possible to obtain final values (for
exponents and cumulants) dilution independent~\cite{DILU3D}.

\begin{figure}
\epsfxsize=0.95\columnwidth
\epsffile{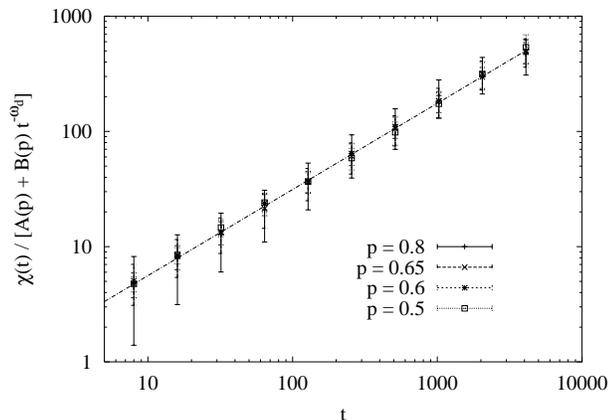}
\caption{The universal part of the susceptibility growth.  The
collapse of the data for different concentrations is the confirmation
that Universality holds.}
\label{fig:asy}
\end{figure}

In order to convince the reader of the goodness of our fits we plot in
Fig.~\ref{fig:asy} the non-connected susceptibility divided by just
the correction-to-scaling factor $[A(p)+B(p)\,t^{-w_d}]$.  If
universality holds ({\it i.e.}\ all the critical exponents, dynamical
and static, are dilution independent) all the data points
(corresponding to four dilution degrees) should collapse on a straight
line in a double logarithm scale. It is clear from this figure that it
is what happens. The equation of the curve is $t^{\gamma \over \nu z}$
with ${\gamma \over \nu z} = 0.748$.

We have shown that it is possible to describe the off-equilibrium
numerical data assuming critical exponents (dynamical as well as
static) independent on the dilution for a wide range of dilutions.
This supports the predictions of the (perturbative) renormalization
group for the statics as well as for the dynamics. So, the
(perturbative) RG scenario that predicts that all the points on the
critical line (for $p<1$) belong to the same Universality class is
very well supported by numerical simulations.

We have found that our estimate of the dynamical critical exponent
$z=2.62(7)$ is incompatible with the experimental value $z=1.7(2)$.
Further numerical and experimental studies should be done in order to
clarify this discrepancy.

We can compare the value of the dynamical critical exponent computed
off- and on-equilibrium. The Heuer's estimate was $z=2.4(1)$ and the
difference with our estimate $z=2.62(7)$ is $z_{\rm off-eq}-z_{\rm
eq}=0.22(12)$, {\it i.e.}\ $1.8$ standard deviations. The conclusion
is that both estimations are compatibles in the error bars.  In any
case, it will be interesting to compute $z$ on-equilibrium by
controlling the scaling corrections.

Moreover, our estimate is not compatible with that of PRG to order
$\sqrt{\epsilon}$ in the $\sqrt{\epsilon}$-expansion ($z=2.34$). The
comparison with the two loops estimates of
$z$~\cite{PRUDNIKOV,JANSSEN} is still worse. 
One possible explanation for this
disagreement could be the lack of Borel summability the diluted model
shows~\cite{NOTE2}. We remark again that the one loop PRG estimates of
the static  critical exponents was very bad (see below).

Another interesting issue is to compare the dynamical
scaling-corrections and the static ones. Unfortunately our statistical
precision is unable to solve this issue. For instance, taking the
values of the $t\ge 4$ we obtain $\omega- \omega_d=0.13(6)$ that is
compatible with zero assuming two standard deviations. If we take the
values of the $t\ge 8$ fit we obtain $\omega- \omega_d=0.35(17)$. We
will devote further work (analytical and numerical) in order to
discern if the leading dynamical scaling-correction corresponds to the
leading static scaling-correction.

We wish to thank H. G. Ballesteros, D. Belanger, L. A. Fern\'andez,
Yu. Holovatch, V. Mart\'{\i}n Mayor and A. Mu\~noz Sudupe for
interesting discussions.

\end{multicols}

\begin{references}
\bibitem{NOTE1} The advantage of the off-equilibrium measurements is
that, as long as $\xi(t)\! \ll\! L$ (where $\xi(t)$ is the
off-equilibrium correlation length), the system size can be considered
infinite for all the practical purposes.
\bibitem{NOTE3} See Ref. ~\cite{CARDY} pag 152. In particular Cardy
shows that to order $\epsilon^\prime$ the eigenvalue associate with
the magnetic field ($y_h$) remains unchanged if dilution is added to
the system. Remember that $y_h=(d+1-\eta)/2$, where $d$ is the
dimensionality of the space.
\end{references}
\end{document}